\documentclass[5p,times,onecolumn,double-spaced]{elsarticle}

\usepackage[utf8]{inputenc}
\usepackage{graphicx}
\usepackage{subcaption}
\usepackage{siunitx}
\usepackage{multirow}
\usepackage{chemformula}
\usepackage{hyperref}
\usepackage{tikz}
    \usetikzlibrary{positioning, shapes, snakes}

\address[thu]{Department of Physics, Tsinghua University, Beijing 100084, China}
\begin{document}

\title{Measurement of the high energy $\gamma$-rays from heavy ion reactions using  \v Cerenkov detector}

\author[thu]{Dawei Si} 

\author[thu]{Yan Zhou} %

\author[thu]{Sheng Xiao} 

\author[thu]{Zhigang Xiao}

\ead{xiaozg@tsinghua.edu.cn}





\begin{abstract}
The energetic bremsstrahlung photons up to 100 MeV produced in heavy ion collisions can be used as a sensitive probe to the short range correlation in atomic nuclei. The energy of the $\gamma$-rays can be measured by collecting the \v Cerenkov light in medium induced by the fast electrons generated in Compton scattering or  electromagnetic shower of the incident $\gamma$ ray. Two types of detectors, based on pure water and lead glass as the sensitive material respectively, are designed for the above purpose. The $\gamma$ response and optical photon propagation in detectors have been simulated based on the  electromagnetic and optical processes in Geant4. The inherent energy resolution of  $0.022+0.51/E_{\gamma}^{1/2}$ for water and $0.002+0.45/E_{\gamma}^{1/2}$ for lead glass are obtained. The geometry size of lead glass and water are optimized at $30\times 30 \times 30$ cm$^3$ and $60\times 60 \times 120$ cm$^3$, respectively, for detecting high energy $\gamma$-rays at 160 MeV. Hough transform method has been applied to reconstruct the direction of the incident $\gamma$-rays, giving the ability to  distinguish experimentally the high-energy $\gamma$ rays produced in the reactions on the target 
from the random background cosmic ray muons.
\end{abstract}


\maketitle

\section{Introduction}\label{sec.I}

Bremsstrahlung high-energy photons produced in heavy ion reactions has attracted increasing interest for its relevance to the nuclear equation of state (nEOS) and to the short range correlation in nuclei. For the studies of nEOS, particularly for the nuclear matter with large neutron-to-proton asymmetry, a variety of isospin probes have been identified  to constrain $E_{\rm sys}(\rho)$ (the density dependent nuclear symmetry energy), including preequilibrium n/p yield ratio\cite{bib:1}, n/p differential flow\cite{bib:2,bib:3} and the Bremsstrahlung high-energy photons \cite{bib:4} etc. Among these probes, Bremsstrahlung $\gamma$-rays created in heavy ion collisions is a clean observable  because of its rare interactions with the medium after it is produced. Very recently, it has been pointed out that the  Bremsstrahlung high-energy $\gamma$ carries the information of the high  momentum tail (HMT) of nucleons, giving rise to the short range correlation effect in nuclei\cite{bib:5,bib:6,bib:7,bib:8}.  On the other hand, however,  the experimental data on this direction is quite scarce.

Recently,  the full $\gamma$ energy spectrum up to 80 MeV has been measured in the reactions $^{86}$Kr+$^{124}$Sn at 25 MeV/u with a 15-unit CsI(Tl) hodoscope mounted on the compact spectrometer for heavy ion experiment (CSHINE)\cite{bib:9,bib:10,bib:11,bib:12}. It has been demonstrated that the $\gamma$ energy spectrum above 20 MeV is reproduced fairly well by the transport model simulations incorporating the $\gamma$ production from the incoherent $np$ scattering with approximate $15\%$ HMT ratio \cite{bib:arxiv}. However, CsI(Tl) is a slow detector, the microsecond response time of CsI(Tl) crystals makes it complicated to reconstruct the total energy from multiple firing units.  Therefore, it is our motivation to develop a fast and relative cheap detector containing sufficiently large-volume sensitive material to detect the high energy $\gamma$-rays in heavy ion reactions. The \v Cerenkov radiation detector is a favorable option because of its fast response time in the order of tens nanoseconds and its ability to infer the incident direction information of the initial $\gamma$-rays, the latter of  which can be used to suppress the cosmic-ray muon background from random directions. 

In this paper, we report the design of a {\v C}erenkov $\gamma$ calorimeter using water and lead glass, respectively, as sensitive medium. Based on Geant 4 packages, the geometry size of the detectors is optimized. The energy resolution is obtained by tracking each \v Cerenkov photon before they arrive at the photomultiplier tube (PMT), of which the quantum response is modeled. The incident direction reconstruction is implemented by Hough transform method.  The paper is organized as following. Section 2 presents the simulations framework of the calorimeter. Section 3 presents the optimization of detector size  and the reconstruction of $\gamma$ direction. Section 4 is the conclusion.

\section{Simulation Setup}  \label{sec. II}
In this study, Geant4 (version 4.10.05)\cite{bib:13} packages are used for Monte Carlo simulation and optimization of the detector. “QBBC” and “G4OpticalPhysics” are applied as the physical process list to describe the electromagnetic (EM) showers of $\gamma$ rays in materials, and to model the generation and transport of {\v C}erenkov photons. For  each event in the simulations, the incident $\gamma$-ray hits the front of the detector. Then the  {\v C}erenkov photons are generated if fast electrons are produced by Compton scattering or by $e^-e^+$ generation. Each  {\v C}erenkov photon is tracked to its termination, either to be absorbed in propagation or to reach the surface of the PMTs, where the waveform pulse of given parameters is generated with a certain quantum efficiency. The  waveform is recorded at an interval of 2 ns for digitization. The final data corresponding to each incident $\gamma$ ray is saved as a matrix of $N_{1} \times N_{2}$ dimensions, where $N_{1}$ represents the number of fired PMTs and $N_{2}$ represents the number of sampling points for the corresponding waveform.

\begin{figure}[!htb]
\includegraphics[width=0.95\hsize]{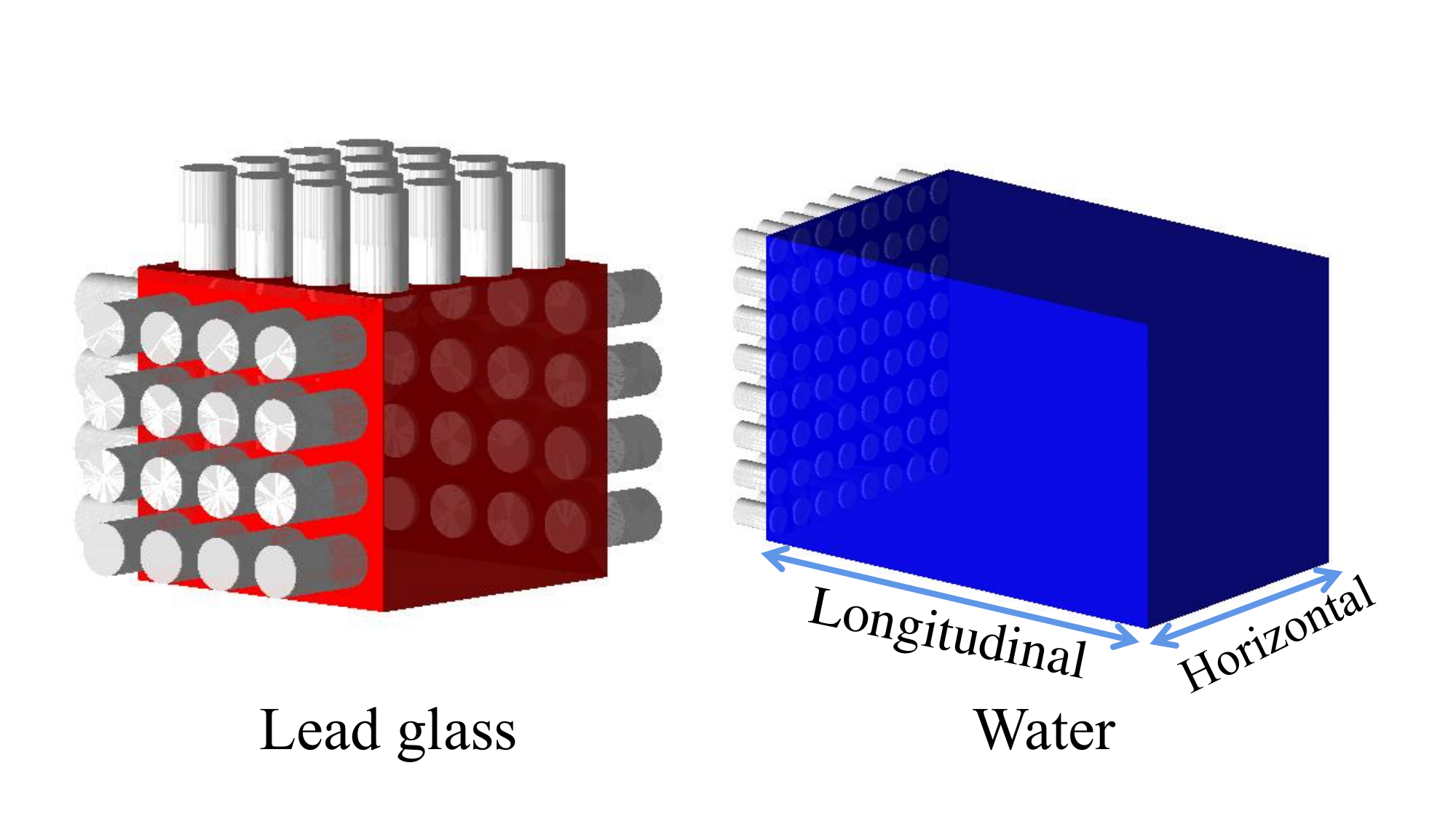}
\caption{(Color online) Detector configuration with two sensitive volumes, lead glass (left) and pure water (right), respectively.}
\label{fig1}
\end{figure}

\subsection{Detector geometry}\label{sec. II1}

The detector structure and the locations of the PMTs are shown in Fig.1 for  lead glass and water as  sensitive material, defined by "G4\_GLASS\_LEAD"(left) and "G4\_WATER"(right) , respectively. The water tank size is 60$\times$60$\times$120 cm$^3$, and the lead glass is 30$\times$30$\times$30 cm$^3$. The PMTs are arranged in an $8\times8$ array in water configuration. For lead glass configuration, PMTs are placed as $4\times4$ arrays on four sides of the detecting tank.  The diameter of the PMT was 51 mm, and the  distance between each neighboring  PMT pair is 70 mm both vertically and horizontally.

\subsection{Optical process}\label{sec. II2}
After invoking the {\v C}erenkov mechanism in Geant4, the energy and number of {\v C}erenkov photons are sampled in each G4step according\cite{bib:14}.

\begin{equation}
\label{eq1}
\frac{d^2N}{d{\lambda}dL}=\frac{2{\pi}{\alpha}Z^{2}}{{\lambda}^{2}}\sin^2{\theta_{\rm c}}
\end{equation}

where $\theta_{\rm c}$ is the {\v C}erenkov angle, $\lambda$ is the wavelength of {\v C}erenkov photon. The initial position of {\v C}erenkov photons is uniformly distributed in every G4step, the emission angle is calculated according to the refractive index of materials and the speed of charged particle, the outgoing azimuth is uniformly distributed within the range of 2$\pi$ , we set a maximum of 100 photons emitted in each step to ensure the detailed sampling. In the process of photon transport, the transmission characteristics of photons in the material and the behavior at the boundary between two materials need to be defined. In this simulation, we defined the scattering and absorption lengths between {\v C}erenkov photons and water molecules by referring to the test data of IceCube\cite{bib:15,bib:16}. Due to the lack of optical parameters of lead glass, we conservatively defined the attenuation efficiency of 70$\%$ for 10 cm propagation. For the boundary characteristics\cite{bib:17,bib:18}, we used UNIFIED model\cite{bib:19,bib:20} in Geant4 and selected "dielectric-dielectric" option to describe the interface between the material and PMTs. In this model, Geant4 will determine the photon boundary behavior according to Fresnel formula and refractive index on both sides. At the remaining boundaries, we used the dielectric\_LUT model\cite{bib:21} and selected the polished Teflon\_LUT boundary. In this way, Geant4 will determine the reflection, refraction and absorption of photons based on built-in parameters.

\subsection{PMT Response}\label{sec. II3}
In the full  situations, photons will be converted into photoelectrons with a certain quantum efficiency after hitting PMT, and pulse will be formed after multiplication. The pulse formed by a single photon is described by \cite{bib:22}
\begin{equation}
\label{eq2}
V_{\rm pulse}(t) =
\begin{cases}
G\exp(-\frac{1}{2}(\frac{t-t_{\rm i}}{\sigma}+e^{\frac{t-t_{\rm i}}{\sigma}})), & t\leq t_{\rm i}\\
G\exp(-\frac{1}{2}(\frac{t-t_{\rm i}}{\sigma})^{0.85}+e^{\frac{t-t_{\rm i}}{\sigma}}), & t\textgreater t_{\rm i}\\
\end{cases} 
\end{equation}
where $t_{\rm i}=t_{\rm hit}+t_{\rm trans}$, $t_{\rm hit}$ represents the time when a photon hits PMT, $t_{\rm trans}=29$ ns represents the electron transit time of PMT, $\sigma=1.2$ ns is the transit time spread. When multiple photons were converted into photonelectrons, the final waveform is generated by  superimposing  all single-photon waveforms as Fig.2(a) shown. Counting from the incidence of  the $\gamma$ ray, the waveform of each PMT within 240 ns was recorded as the final data. Fig.2(b) shows the distribution of the time when optical photons reach on PMTs in lead glass configuration which was extracted by linear fitting the rising edge of waveform\cite{bib:23}. It illustrates most photons reach on the surface of PMTs between 26ns and 31ns after the $\gamma$ emission, which means we can distinguish between direct and scattered photons according to the distribution in the direction reconstruction (see section 3). 
\begin{figure}[htb]
\includegraphics[width=\hsize]{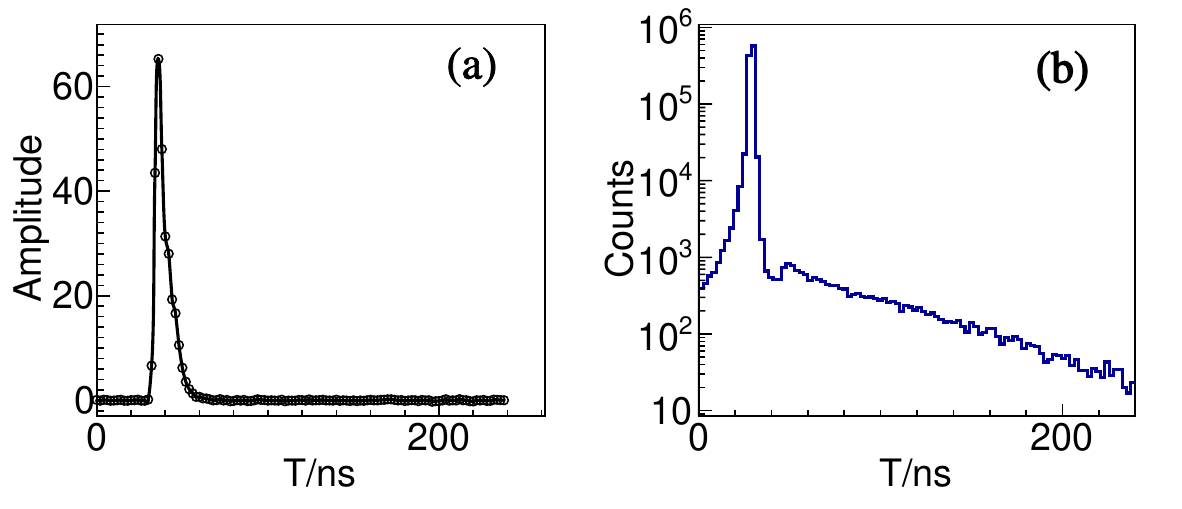}
\caption{(Color online) (a) a typical waveform for PMT, (b) the distribution of the time when optical photons reach on PMTs in lead glass detector.}
\label{fig2}
\end{figure}

\begin{figure}[htb]
\includegraphics[width=\hsize]{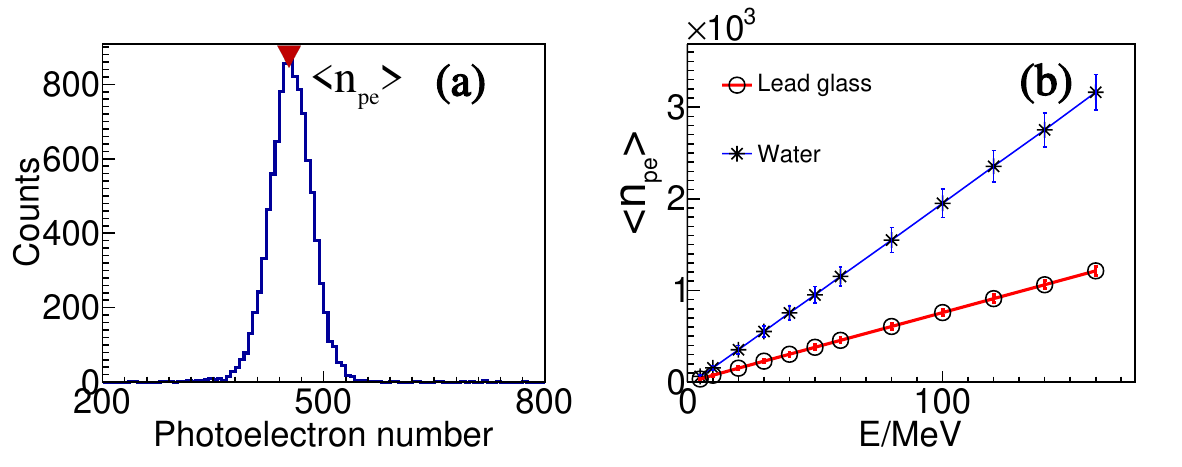}
\caption{(Color online) (a)the spectrum of the photoelectron yield  for 50 MeV $\gamma$ rays in lead glass detector,(b) Linear response of the calorimeter. }
\label{fig3}
\end{figure}

\section{Result and discussion}  \label{sec. III}
\subsection{Influence of detector size on energy resolution}\label{sec. III1}
We use the the photoelectron peak number $\left<n_{\rm pe}\right>$  and the energy resolution, defined by $\delta_{E_\gamma}=\delta_{n_{\rm pe}}=\sigma_{ n_{\rm pe}}/\left<n_{\rm pe}\right>$,  to optimize the detector design. In the simulation, such high-energy $\gamma$ rays hit perpendicularly the  center of the front surface of detector. The shower electrons and positrons, if produced with velocity exceeding the speed of light in the medium,  will generate {\v C}erenkov light propagating to the PMT where the photoelectrons are generated. Due to statistic fluctuations, the number of photoelectrons varies. Fig. 3 (a) presents the distribution of the number of photoelectrons for 50 MeV incident $\gamma$ ray in the lead glass detector as an example. The photoelectron peak number  $\left<n_{\rm pe}\right>$ is taken as the average number of photoelectrons in the following analysis.  Fig. 3 (b) presents the distribution of   $\left<n_{\rm pe}\right>$  as a function of incident $\gamma$ energy  $E_\gamma$ for the two configurations at their own optimized volume (see below). It is shown that for either detector with given size,   $\left<n_{\rm pe}\right>$  exhibits a linear dependence on $E_\gamma$. Thus, the $\gamma$ ray energy can be measured by the number of photoelectrons.
\begin{figure}[htb]
\includegraphics[width=\hsize]{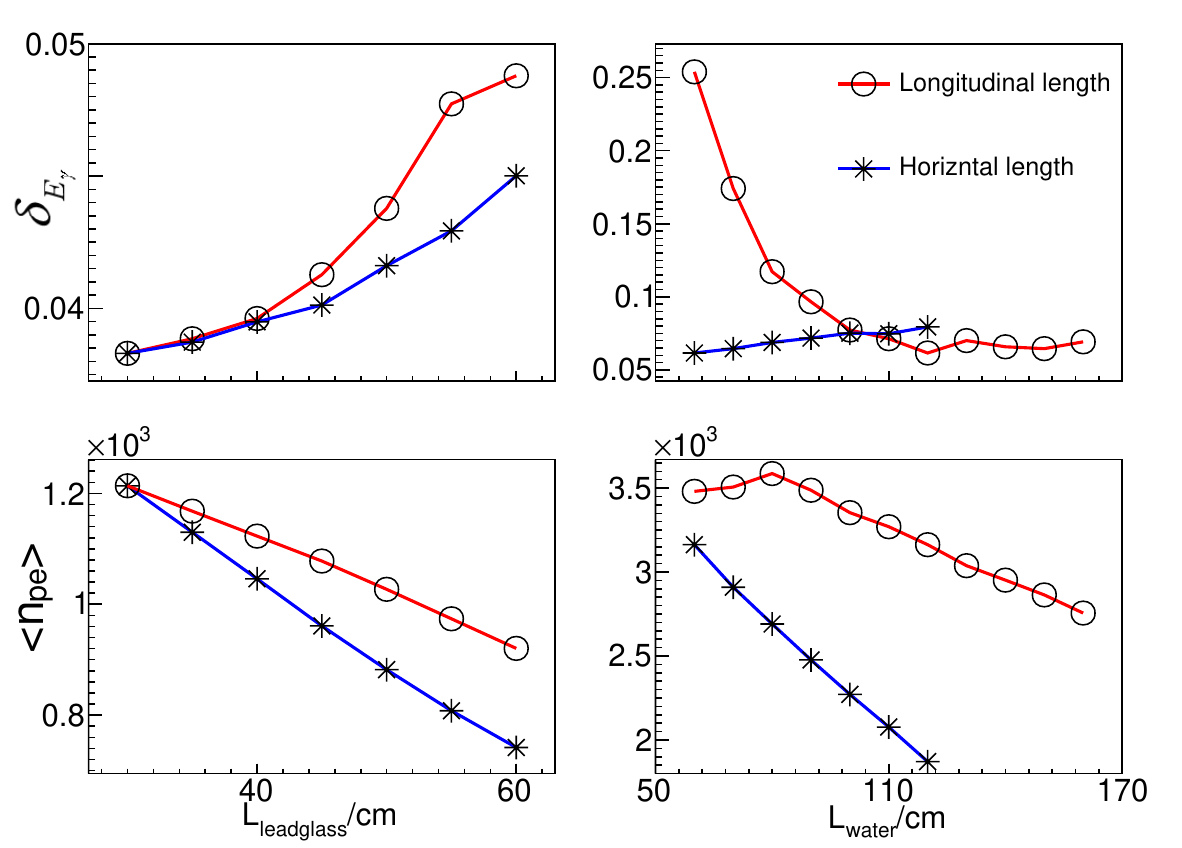}
\caption{(Color online) process of size optimization.(a)relationship between energy resolution and detector size for lead glass, (b)relationship between energy resolution and detector size for water, (c) relationship between $\left<n_{\rm pe}\right>$ and detector size for lead glass, (d)relationship between $\left<n_{\rm pe}\right>$  and detector size for water. }
\label{fig.4}
\end{figure}

We then optimize the detector size  at a given maximum $\gamma$ energy of 160 MeV, which covers the range of interest of $E_\gamma$ in the heavy ion reactions at Fermi energies.   In each event the distribution of photoelectron number is analyzed  to obtain the  $\left<n_{\rm pe}\right>$ and the standard deviation ($\sigma_{n_{\rm pe}}$) of the distribution.  If the size of the detector medium is too small, much  of the $\gamma$ ray  energy will leak to outside of the sensitive volume. On the other hand,  if the detector medium is too large, {\v C}erenkov photon will be scattered many times and gradually absorbed, leading to the reduction of photoelectron number collected by PMTs. These two factors compete with each other and determines the energy resolution.  Fig.4(a) and (c) illustrate the energy resolution  $\delta_{E_\gamma}$ and $\left<n_{\rm pe}\right>$  as a function of the horizontal and longitudinal length for the lead glass configuration. Clearly, as the horizontal and longitudinal length increases, $\delta_{E_\gamma}$ increases, while and  $\left<n_{\rm pe}\right>$ decreases, respectively. So $30 \times 30 \times 30$ cm$^3$ is the optimal size for lead glass.  Fig.4(b)(d) shows the same quantities in pure water configuration. With increasing the longitudinal length, the energy resolution  $\delta_{E_\gamma}$  decreases gradually and converges to 6\%,   and $\left<n_{\rm pe}\right>$  first increases and then decreases, because {\v C}erenkov light attenuation becomes the main factor after the longitudinal length exceeds 80cm. As horizontal length increases, $\delta_{E_\gamma}$ increases and  $\left<n_{\rm pe}\right>$  decreases, respectively. So $60 \times 60 \times 120$ cm$^3$ is the optimal size for water.

\begin{figure}[htb]
\includegraphics[width=\hsize]{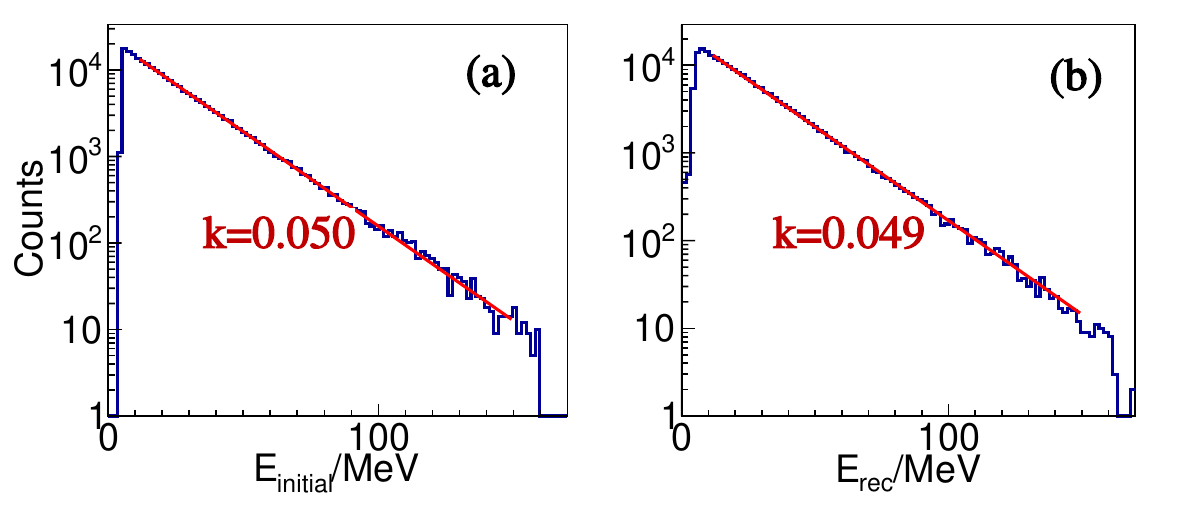}
\caption{(Color online) The initial(a) and reconstructed(b)  $\gamma$ energy spectra in lead glass configuration.}
\label{fig.5}
\end{figure}

\begin{figure}[htb]
\includegraphics[width=\hsize]{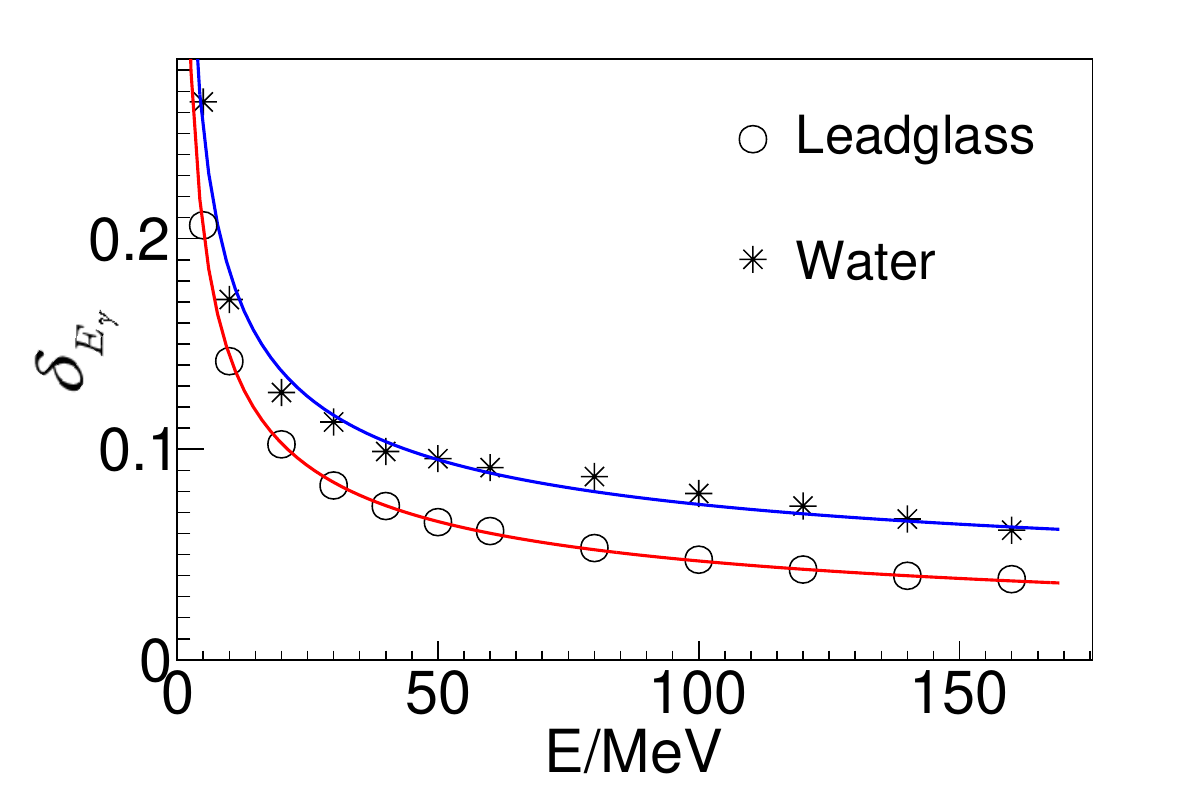}
\caption{(Color online) Resolution prediction of the calorimeter of water and lead glass, respectively.}
\label{fig.6}
\end{figure}

Given the good linear response for the water and lead glass {\v C}erenkov calorimeter to the $\gamma$ energy, as shown in Fig. 3 (b), one can reconstruct the $\gamma$ energy from the signal height equivalent to the number of photoelectrons. To test the ability, we simulate the detector response for $10^5$ $\gamma$ events with the initial energy $E_{\rm initial}$ in an exponential distribution.   The slope of the input exponential  distribution is set as $-0.05$, as shown in Fig. 5 (a). The reconstructed  energy ($E_{\rm rec}$)  is plotted in panel (b), with the slope parameter being fitted at $-0.049$. It is shown that the  {\v C}erenkov calorimeter of lead glass measures the high-energy $\gamma$ in the range of 5-160MeV. Fig. 6 shows the resolution at various incident energies for lead glass and water configurations, respectively.  The inherent resolution of $0.022+0.51/E_{\gamma}^{1/2}$ for water and $0.002+0.45/E_{\gamma}^{1/2}$ for lead glass are obtained by fitting the simulated data points. At high energies(above 100 MeV), the resolutions saturate at about $7.3\%$ and $4.7\%$, respectively. 

\subsection{Direction reconstruction}\label{sec. III2}
As well known, there is a definite angle between  {\v C}erenkov photons and the charged particle\cite{bib:14} which is the basis for direction reconstruction. In fact, $\gamma$ shower also partially retain this feature, Fig.7 shows the angle distribution between the {\v C}erenkov radiation direction and the initial direction of electrons ($\gamma$ rays) in water (lead glass) which was obtained by Geant4 simulation, the energy of electron and $\gamma$ was sampled evenly from 5 to 160MeV in the simulation. The refractive index of lead glass and water are 1.7 and 1.3, so the cosine of their {\v C}erenkov angle are $\cos\theta_{\rm c}\approx 0.58$ and 0.77, respectively. According to Fig.7, although the $\rm e^+e^-$ pair production  and Compton effect may cause scattering, the emission angle distribution of {\v C}erenkov photons produced by the EM shower is still related to the initial direction of $\gamma$ rays. This indicates that the direction of $\gamma$ rays can be reconstructed by referring to the method of electron direction reconstruction used in large experiments such as Super-Kamiokande and Sudbury Neutrino Observatory(SNO)\cite{bib:25,bib:26}. It is found in our work that the {\v C}erenkov photons experience scattering  many times before reaching PMTs in water due to the overlength of the medium, smearing heavily the initial direction information, so we only reconstructed the $\gamma$ rays direction in the lead glass configuration. 

\begin{figure}[htb]
\includegraphics[width=\hsize]{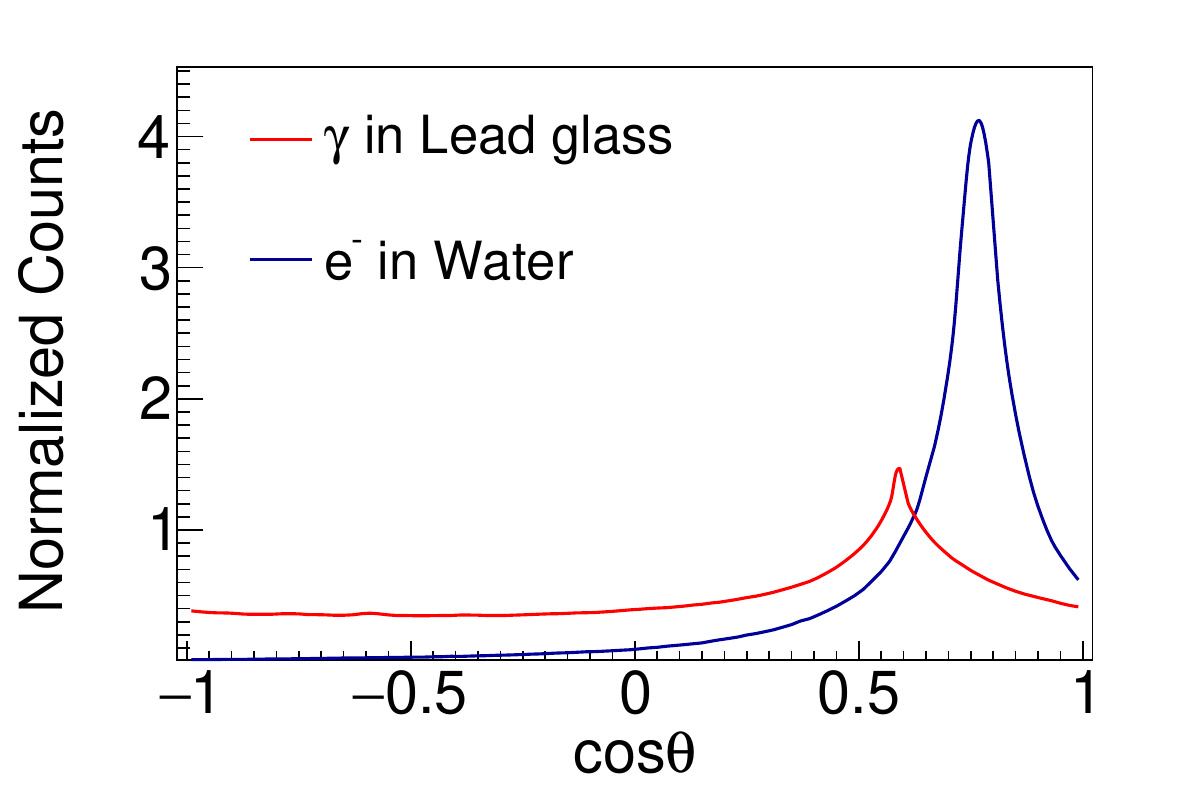}
\caption{(Color online) {\v C}erenkov photon direction distribution for electron incidence in water and $\gamma$ ray incidence in lead glass, respectively.}
\label{fig.7}
\end{figure}

\subsubsection{Vertex reconstruction}\label{sec. III2}
For reconstructing the direction of electrons, it is usually assumed that electrons emit {\v C}erenkov light from a fixed point. According to the angle distribution in Fig. 7, it can be assumed that the $\gamma$ rays emit {\v C}erenkov light from a fixed point with a specific {\v C}erenkov angle, and hence the time of photon reaching the PMT can be expressed as \cite{bib:27,bib:28}

\begin{equation}
\label{eq3}
t_{\rm hit}=\frac{\mid \vec{X}_{\rm pmt}-\vec{X}_{\rm vtx} \mid}{v}+t_{0}
\end{equation}
where $t_{0}$ represents the moment when the {\v C}erenkov light is generated, $t_{\rm hit}$ represents the moment when the photon hits PMTs, $v$ is the velocity of light in lead glass, $\vec{X}_{\rm pmt}$ and $\vec{X}_{\rm vtx}$ are the coordinate of the PMT and the vertex respectively.In our analysis, the optimal estimation of the vertex coordinates is obtained by minimizing the $\chi^2$ of fitting the time distribution with formula (3), in which the $t_{0}$ and $\vec{X}_{\rm vtx}$ are fitting parameters. In each $\gamma$ event, the timing on each PMT is extracted by linear fitting to the rising edge of waveform, where the crossing point of the linear fitting and zero baseline is taken as the timing signal of the PMTs\cite{bib:23}. The spatial coordinate of each firing PMT is used as $\vec{X}_{\rm pmt}$. Since the reflector layer is set in the simulation, some {\v C}erenkov photons will be reflected before hitting the PMTs  and the timing signals will deviate from formula (3). So according to Fig. 2 (b), we only selected the PMTs with the hit time being less than 1.5 ns before the peak and 1 ns after the peak of the time distribution. We defined the coordinate for the center of lead glass as ${\rm (0 ~cm,0 ~cm,0 ~cm)}$.  Fig. 8  shows the $\chi^2$ distribution the vertex coordinate fitting for a 10 MeV $\gamma$ ray incidence, indicating the optimal vertex coordinate at ${\rm (1.29~cm,-1.11 ~cm,-6.34 ~cm)}$.
\begin{figure}[htb]
\includegraphics[width=\hsize]{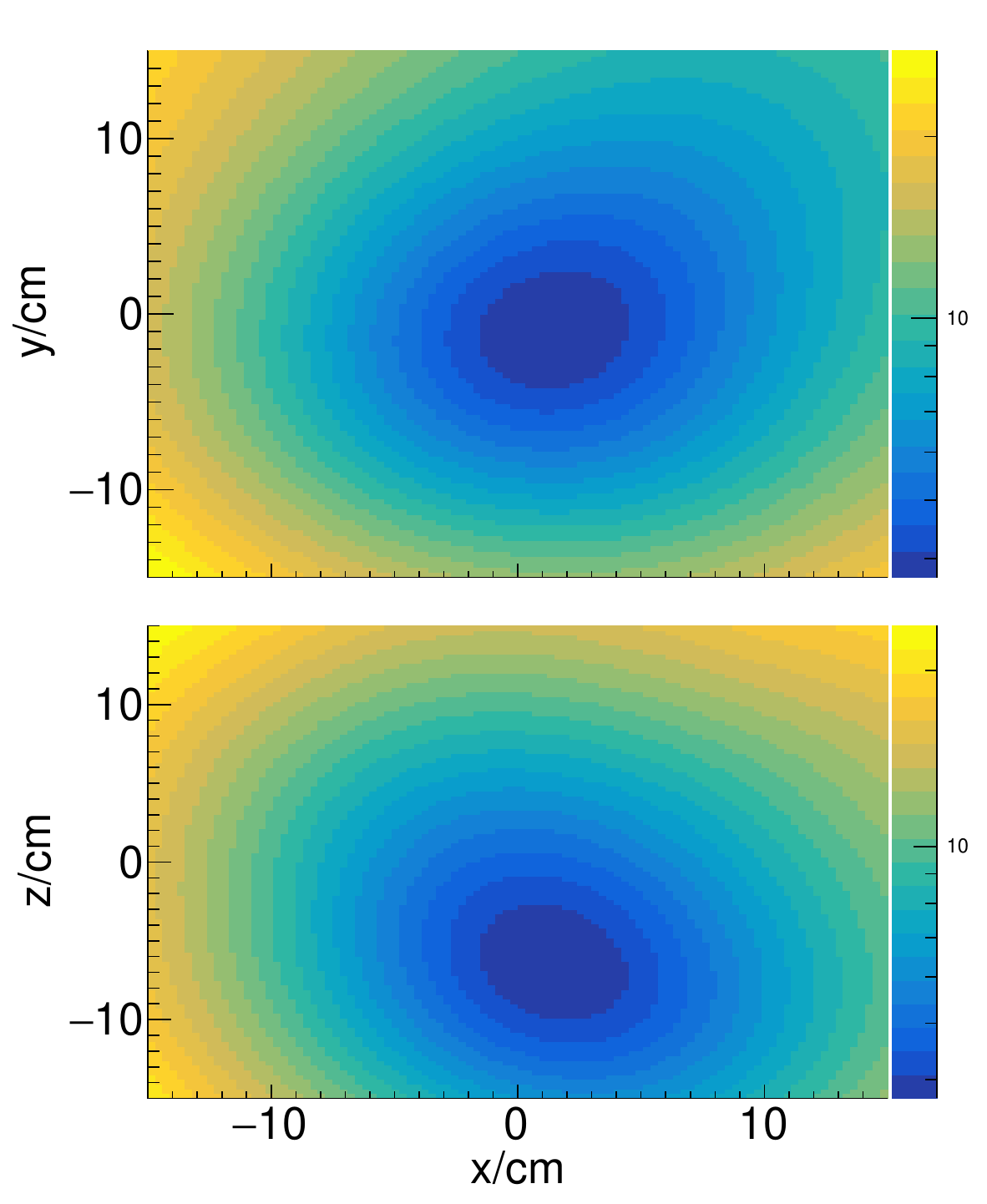}
\caption{(Color online) The $\chi^2$  distribution contour in the coordinate space of the vertex fitting.}
\label{fig.8}
\end{figure}

\begin{figure}[htb]
\centering
\includegraphics[width=0.8\hsize]{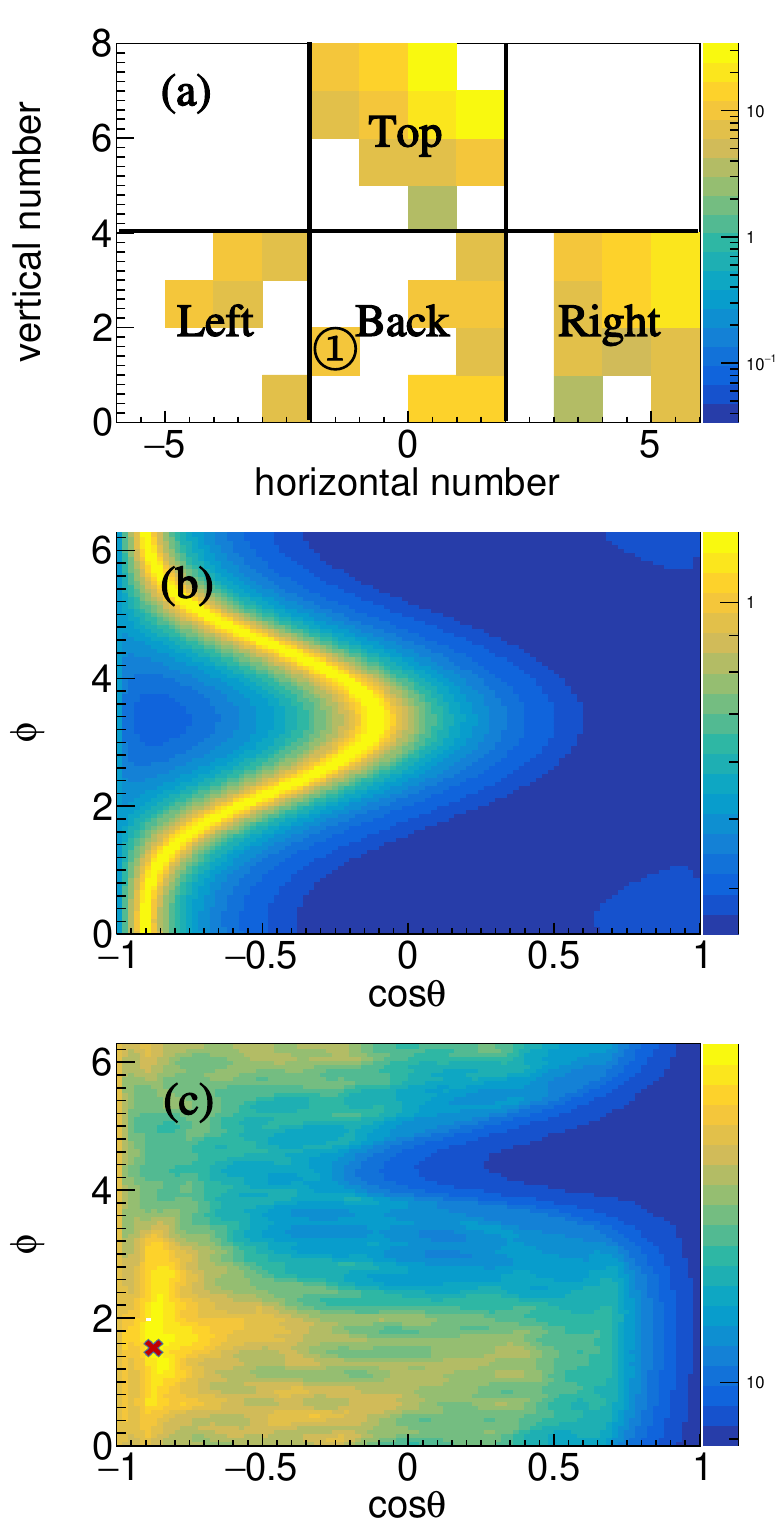}
\caption{(Color online) Event display of Hough transformation. (a) is the position distribution of firing PMTs, there are four sides to place PMTs in lead glass configuration, top, left, right and back, (b) is the result of Hough tranform for the marked $ 1^{\rm st}$ PMT on the back surface, (c) is the cumulative result of Hough transform for all firing PMTs in the time window. The cross indicates the optimized direction.}
\label{fig.9}
\end{figure}
\subsubsection{Hough transform}\label{sec. III2}
Hough transform\cite{bib:29,bib:30,bib:31} has been successfully applied  to identify {\v C}erenkov rings, which can  map the vector space of the vertex-to-PMT direction  to the vector space of the electron incident direction. An example of such application can be found in Super-Kamiokande\cite{bib:32}. Similarly, we can define the vector from the vertex of the $\gamma$ ray to the firing PMT and the initial direction vector of the $\gamma$ ray as $\vec{V}_{p}$ and $\vec{V}_{\gamma}$ respectively, and $\theta$ represents the angle between these two vectors. The probability distribution of $\theta$ is shown as the red line in Fig.7. The vector space of incident direction of $\gamma$ ray was divided by $100 \times 100$ according to ($\cos\theta$,$\phi$), and the weight of each cell can be expressed as
\begin{equation}
\label{eq4}
    W_{\rm ij}=\sum_{1}^{k}f(\cos\theta_{\rm ijk}),\quad \cos\theta_{\rm ijk}=\vec{V}_{{\gamma}\rm ij}\cdot\vec{V}_{\rm pk}
\end{equation}
$\vec{V}_{{\gamma}\rm ij}$ represents the central unit vector of the cell at row $i$ and column $j$ in the vector space of incident direction of $\gamma$ ray, $\vec{V}_{\rm pk}$ represents the unit vector from the vertex pointing to the center of $k^{\rm th}$ firing PMT, and function $f$ represents the {\v C}erenkov angle distribution function of $\gamma$ rays (the red line in Fig.7) in lead glass. Fig.9 shows the event display of Hough transform for the incident $\gamma$ event in the direction of $(0.068, 0.063, -0.995)$. Fig.9(a) shows the hitting PMTS position distribution for this event, the color represents the signal amplitude in the corresponding PMT; Fig.9(b) shows the result of hough transform for the $1^{\rm st}$ PMT; Fig.9(c) shows the cumulative result of Hough transform for all the firing PMTs, the brightest point in Fig.9(c) represents the optimal estimate of the incident gamma direction. The optimal estimate is $(-0.077,0.487,-0.87)$, and the deviation from the initial incidence direction is $26.9^\circ$ for this event as an example.

\begin{figure}[htb]
\includegraphics[width=\hsize]{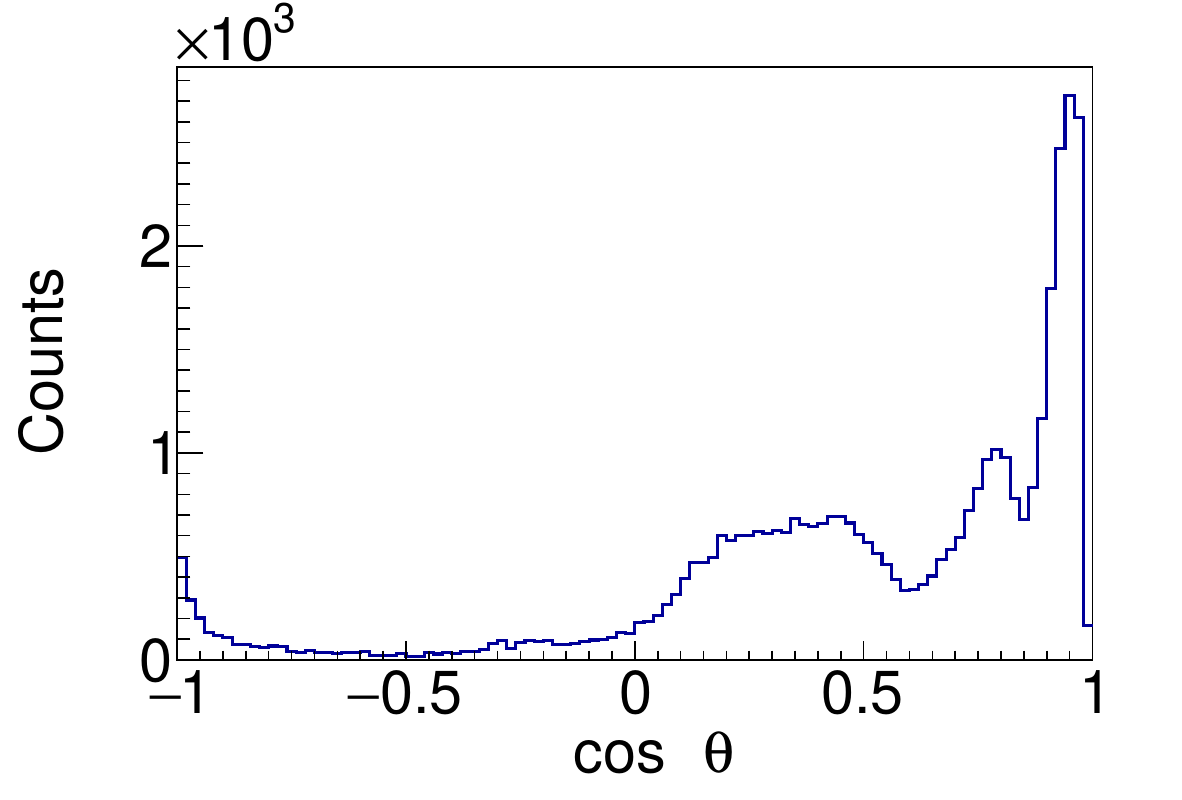}
\caption{(Color online) The distribution of the angle between initial and reconstructed direction of $\gamma$ rays}
\label{fig.10}
\end{figure}

Fig.10 shows the  distribution of $\cos\Delta\theta$ ,where $\Delta\theta$ is the angle between the reconstructed direction and the initial direction of $\gamma$ rays hitting  the front of lead glass uniformly from the target. It is shown that the peak of the cosine values is near $\cos\Delta\theta=1$, indicating that the detector is able to reconstruct the direction of signal in the lead glass configuration. But the cosine distribution is broadened  considerably due to  the rough assumption that the $\gamma$ rays emit {\v C}erenkov light at a fixed point. In fact, according to the red line in Fig.7, most $\gamma$ rays would generate {\v C}erenkov light in a path whose length is comparable to the detector size, which is contribute to the bias in direction reconstruction. And the anti-symmetry of the locations of the  PMTs also cause  the bias.

\begin{figure}[htb]
\includegraphics[width=\hsize]{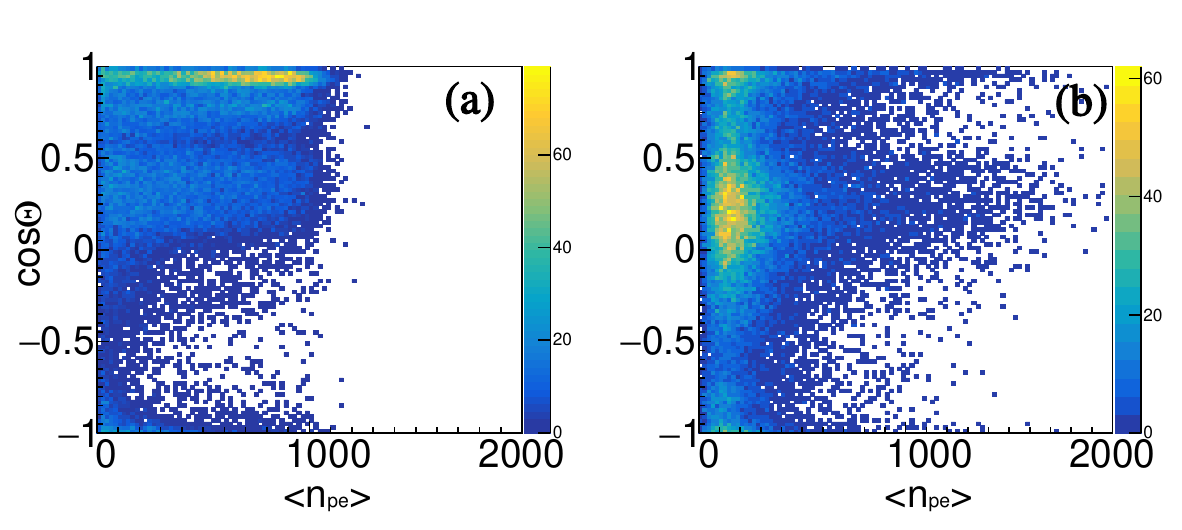}
\caption{(Color online) Two-dimensional distribution of $\cos\Theta$ and $\left<n_{\rm pe}\right>$ for $\gamma$ rays(a) from the reaction target and for cosmic ray muons (b).}
\label{fig.11}
\end{figure}

\subsubsection{Discrimination between $\gamma$ and cosmic ray muon}\label{sec. III2}
In real beam experiment, only the $\gamma$ rays from the reactions on the target are of interest. Since the direction of $\gamma$ rays from the reaction target is different from cosmic ray, it provides a way to suppress the background. To test the ability of suppressing the cosmic-ray background, we generate $\gamma$ rays with energy between  $5-160$ MeV to the front of detector and mix with uniform $\mu^{-}$ emissions from  the top of detector. The  $\mu^{-}$ energy  $E_{\mu}$ (in GeV) and zenith angle $\theta_{\mu}$ was sampled using the Gaisser formula \cite{bib:33} 
\begin{equation}
\label{eq5}
    \frac{dI}{dE_{\mu}d\cos\theta_{\mu}}=\frac{0.14}{E_{\mu}^{2.7}}\left[\frac{1}{1+\frac{1.1E_{\mu}\cos\theta_{\mu}}{ 115}}+\frac{0.054}{1+\frac{1.1E_{\mu}\cos\theta_{\mu}}{850}}\right]
\end{equation}
Considering that the threshold for $\mu^{-}$ to produce {\v C}erenkov in lead glass is 78 MeV, we set the sampling range of $80-1000$ MeV. Here the physical quantity $\Theta$ denotes  the angle between the reconstructed direction and the vector from reaction target to the fitted vertex vector. Fig.11 shows the two-dimensional distribution of cos$\Theta$ and  $\left<n_{\rm pe}\right>$ for $\gamma$ rays from target (a) and for cosmic rays (b), respectively. Very different feature between the reaction $\gamma$ rays and the cosmic ray muon background is evident.  The $\cos\Theta$ of  $\gamma$ rays is concentrated above 0.5 and the  $\left<n_{\rm pe}\right>$ is relatively evenly distributed between 0 and 1000, while the cos$\Theta$ of $\mu^{-}$ is concentrated between 0 and 0.5 and the  $\left<n_{\rm pe}\right>$  is concentrated around 200.  Therefore, the directivity of the {\v C}erenkov light provides a new dimension information for distinguishing the signal from the background.
 
\section{Conclusion}  \label{sec. IV}

In this work, we investigate the feasibility of using {\v C}erenkov  calorimeter to detect the bremsstrahlung $\gamma$ rays from heavy ion reactions at Fermi energies.  A full framework has been established to simulate the response and performance of the {\v C}erenkov gamma calorimeter based on Geant 4 packages, including $\gamma$ induced EM shower, {\v C}erenkov photon generation and propagation, and the parameterization of PMT waveform. The optimal volume, linear response and energy resolution of the detector are obtained with water and lead glass being the sensitive medium, respectively. The inherent energy resolutions at $0.022+0.51/E_{\gamma}^{1/2}$ level for water and $0.002+0.45/E_{\gamma}^{1/2}$ for lead glass are predicted. It is demonstrated that the initial direction of $\gamma$ rays can be reconstructed by using vertex fit and Hough transform method, showing the ability to distinguish the bremsstrahlung $\gamma$ rays produced in the  reactions from  the cosmic ray muon background  in real beam experiment. The detector will be built  and applied shortly to measure high-energy $\gamma$ rays produced in heavy ion reactions.

\section* {Acknowledgements}
Supported  by the Ministry of Science and Technology under Grant Nos. 2022YFE0103400 and 2020YFE0202001,by the National Natural Science Foundation of China under Grant Nos. 11961141004, 12205160 
 and by Tsinghua University Initiative Scientific Research Program.

\end{document}